\documentclass[review]{elsarticle}

\usepackage{hyperref}

\newcommand{\be}{\begin{equation}}
\newcommand{\ee}{\end{equation}}

\begin{document}
\def\balpha{\mbox {\boldmath ${\alpha}$}}
\def\chat{{{\bf {\hat c}}}}
\def\hhat{{{\bf {\hat h}}}}

\begin{frontmatter}

\title{Primitive models of room temperature ionic liquids. Liquid-gas phase coexistence}

\author[mymainaddress]{Y. V. Kalyuzhnyi\corref{mycorrespondingauthor}}
\author[mysecondaryaddress]{J. Re\v{s}\v{c}i\v{c}} 
\author[mymainaddress]{M. Holovko} 
\author[mytertiaryaddress]{P. T. Cummings} 

\cortext[mycorrespondingauthor]{Corresponding author}

\address[mymainaddress]{Institute for Condensed Matter Physics NASU, Lviv, Ukraine}
\address[mysecondaryaddress]{Faculty of Chemistry and Chemical Technology, University of Lubljana, Slovenia}
\address[mytertiaryaddress]{Department of Chemical and Biochemical Engineering, Vanderbilt University, Nashville,
Tennessee 37235-1604, USA}
\begin{abstract}

We propose several versions of primitive models of room temperature ionic liquids 
(RTILs) and develop a mean spherical approximation (MSA)-type theory for their 
description. RTIL is modeled as a two-component mixture of hard-sphere anions
and flexible linear chain cations, represented by tangentially bonded hard
spheres with the charge located on one of the terminal beads. The theoretical description of
the model is carried out using the solution of the appropriately modified associative MSA (AMSA). Our solution reduces to solving one nonlinear algebraic equation for the 
Blum's screening parameter $\Gamma$, which in turn is used to express all thermodynamic 
properties of the models of interest. We calculate liquid-gas phase diagrams using theoretical and computer simulation methods for two versions of the model, represented by the dimer ($D$) 
and chain ($C$) models. Theoretical predictions for the phase diagrams appear to
be in reasonably good agreement with computer simulation results. 
It is demonstrated that the models and theory are able to qualitatively reproduce 
experimentally observed phase behavior of RTILs, in particular the decrease of the critical 
temperature and critical density with increasing asymmetry of the model in its 
shape and position of the charge.

\end{abstract}


\end{frontmatter}


\section{Introduction}

Much of success achieved in the statistical-mechanical theory of liquids is due to the
availability of models that are simple enough to be described analytically
and at the same time are able to reproduce the most important features of a targeted 
class of systems. In the case of electrolyte solutions, molten salts and liquid metals 
the models of this type are represented by the so-called 'primitive' models (PMs). These are 
the models, which combine short-range hard-sphere repulsion and long-range Coulomb 
interaction. Sufficiently accurate and simple theoretical description of these models can be 
achieved in the framework of the mean spherical approximation (MSA). An important advantage of 
the MSA is that for most of the versions of PMs it can be solved 
analytically, yielding relatively simple expressions for the structure and thermodynamic 
properties of the system. Waisman and Lebowitz derived an analytic solution of the MSA for restricted 
PM (RPM) of electrolytes \cite{Waisman1972a,Waisman1972b} (equivalent two-component mixture 
of equal size charged hard spheres) using Laplace transform techniques. This solution was 
elaborated and extended to the general case of any number of components with arbitrary charges (subject to electroneutrality) and hard-sphere diameters by Blum \cite{Blum1975}. He was able 
to reduce the problem to the solution of only one nonlinear algebraic equation for 
the famous Blum's scaling (screening) parameter $\Gamma$, which appears to be the MSA analogue
of Debye screening length. All MSA thermodynamic properties can be expressed in terms of this
parameter \cite{Blum1977}.

The major goal of this study is to propose the extension of the PMs of electrolyte solutions
for room temperature ionic liquids (RTILs) and develop their theoretical 
description. Although the vast majority of the previous studies have been focused on the 
description of a RTIL on the detailed atomistic level (see, e.g.\cite{real_review}), 
over the last decade several simple 
models of RTIL have been proposed and investigated
\cite{Malvaldi2007,Spohr2009,Martin2009,Camp2011,Lindergerg2014,Linderberg2015,Alcantara2016,Forsman2016}. Molecular ions (cations) in the framework of these models are represented either as a 
Lennard-Jones/hard spheres with off-center point charges
\cite{Spohr2009,Camp2011,Lindergerg2014,Linderberg2015,Forsman2016}, as a 
Lennard-Jones/hard-sphere dimers with point charges located on one or both sites 
\cite{Malvaldi2007,Camp2011,Alcantara2016} or as a hard spherocylinder with point charge 
located on one of its ends \cite{Martin2009}. In these papers the structural and dynamic
properties \cite{Malvaldi2007,Spohr2009,Camp2011,Lindergerg2014,Linderberg2015,Forsman2016},
density and potential profiles in a planar electrical double layer \cite{Alcantara2016},
 as well as liquid-solid \cite{Lindergerg2014,Linderberg2015,Forsman2016}
and gas-liquid \cite{Martin2009,Camp2011} phase behavior have been studied.
The common feature of all these studies is that the properties of the models were investigated using computer simulation methods only, either Molecular Dynamics (MD) 
or Monte Carlo (MC) simulations. 

We propose to model an RTIL as a two-component mixture of hard-sphere
anions and flexible linear chain cations, represented by the tangentially bonded hard
spheres with the charge located on one of the terminal beads. For a cation chain length
of two, our model reduces to the model studied earlier \cite{Alcantara2016}. More
important is that in addition we propose the MSA type of the theory, which is able to provide an analytical description of the structural and thermodynamic properties of the model. The
theory is based on the multidensity version of the MSA \cite{Wertheim1984,Wertheim1986,amsa},
the so-called associative MSA (AMSA), and represents its extension for chain-forming fluids 
\cite{Protsykevytch,Kalyuzhnyi2001}. We are focused here on the liquid-gas phase behavior of 
the model, which appears to be one of the most important characteristics of the RTILs. 
Prediction of the phase diagram and critical parameters for the PM of electrolytes has long been a challenge for the theory and computer simulation since the late sixties and early
seventies, when its existence for the RPM had been suggested both theoretically 
\cite{Stillinger1968,Ebeling1971,Stell1976} and via computer simulation \cite{Vorontsov1970}
(for more detailed historical review see Refs. \cite{Stell1995,Weingartner2001,Hynninen2008}). Due to a number of computer simulation studies, the exact position of the critical point and
the phase diagram of the RPM and PMs with different hard-sphere sizes and charge ratios have been recently identified \cite{caillol,panagiotop2002,panagiotop2003}. 
The situation with theoretical estimates
is less satisfactory, i.e. while the position of the critical point is predicted with 
reasonable accuracy the phase diagram is still too narrow in comparison with computer 
simulation phase diagrams
\cite{Fisher1993,Fisher1994,Levin1996,Patsahan2006,Patsahan2010a,Patsahan2010b}. 
Perhaps the most accurate theoretical results for the RPM were obtained by Blum and co-workers \cite{Jiang2002}. They assumed that due to strong Coulomb interaction cations and
anions form dimers and at low enough temperatures the system is completely dimerized.
For their theoretical description, a combination of the usual MSA and AMSA was developed and applied.
We note in passing that in this paper the authors refer to AMSA as to binding MSA, although
both are identical.
The theory yields fairly accurate description of the critical temperature and critical density
when compared with computer simulation. The scheme suggested by Blum and co-workers was extended
for size asymmetric PMs by Qin and Prausnitz \cite{Qin2004}. Recently this scheme was used
to describe phase behavior of the PMs confined in a disordered porous media by 
Holovko et al. \cite{Holovko2017a,Holovko2017b}. In our study we propose further
extension of the Blum's approach and apply it to primitive models of RTILs.

The remainder of the paper is organized as follows. In Section 2 we introduce the
model and in Section 3 we discuss the AMSA theory. Solutions of the AMSA for the general 
version of the model and for two simplified versions are derived in Section 4 and in
Section 5 we present expressions for the structure and thermodynamic properties of these 
models. In Section 6 we consider an extension of Blum's approach to calculate the phase 
diagrams and in Section 7 we discuss details of computer simulation approach. Our
results and discussion are presented in Section 8 and in Section 9 we collect our conclusions

\section{The model}

\begin{figure*} 
\begin{center}
\includegraphics[width = 0.9\textwidth]{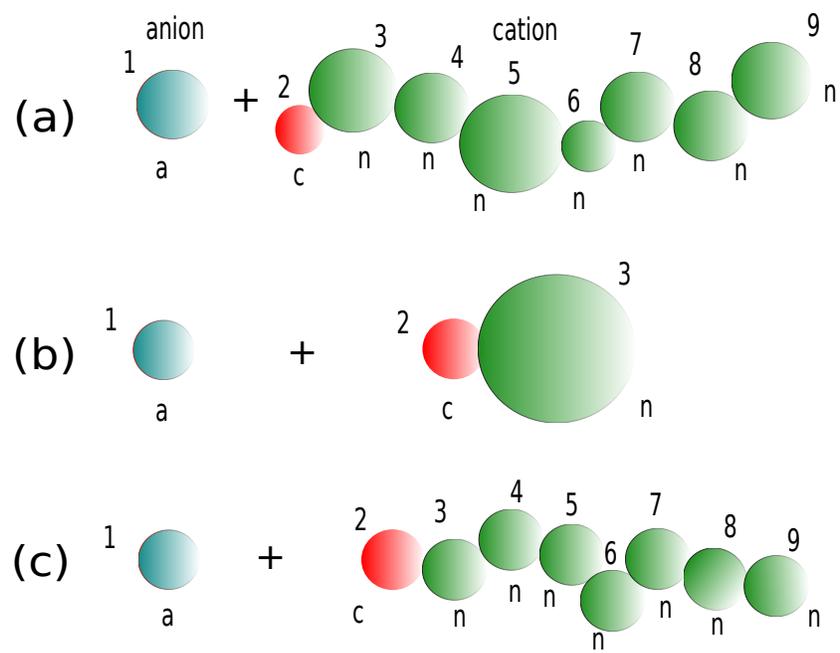}
\caption{Schematic representation of the primitive models of the RTIL. The general version
of the model (a) and its dimer (b) and chain (c) versions, $D$ and $C$, respectively.}
\label{fig1}
\end{center}
\end{figure*}

We are modeling an RTIL as a two-component mixture of hard-sphere anions with the number 
density $\rho_a$ and flexible linear chain cations with the number density
$\rho_c$, represented by $m-1$ tangentially bonded hard
spheres with the charge located on one of the terminal beads.  
The pair potential acting between the particles is represented by the sum of 
site-site hard-sphere potentials $U^{(hs)}_{ij}(r)$ :
\be
U_{ij}^{(hs)}(r)  = \left\{
\begin{array}{rl}
\infty, & r<\sigma_{ij}=(\sigma_i+\sigma_j)/2, \\
0, & r>\sigma_{ij},
\end{array}
\right.
\label{hs}
\ee
and Coulomb potential $U^{(C)}_{ij}$:
\be
U_{ij}^{(C)}(r)=
(\delta_{i1}\delta_{j2}+\delta_{i2}\delta_{j1}){e^2z_iz_j\over 4\pi
\epsilon\epsilon_0 r},
\label{C}
\ee
which is valid between the charged hard-sphere sites. Here $\sigma_i$ is the size
of the hard-sphere site $i$,
$\delta_{ij}$ is the Kronecker delta,
 $\epsilon$ and $\epsilon_0$ are the dielectric constants of the continuum and vacuum, respectively, 
$ez_i$ is the charge of the site $i$
and the site species indices $i,j$ are taking the values $1,2,\ldots,m$, with $i=1,2$ 
denoting anion and charged bead of the cation, respectively (see figure \ref{fig1}).
We assume also that $|z_1|=|z_2|=z$ and the total number density of the system is
$\rho_{t}=2\rho$, where $\rho=\rho_a=\rho_c$.

\section{Theory}

The thermodynamic properties of the model are derived using an appropriately
modified Wertheim's multidensity Orstein-Zernike (OZ) equation supplemented by the
associative mean spherical approximation (AMSA) \cite{amsa} formulated for  
chain-forming fluids \cite{Protsykevytch,Kalyuzhnyi2001}. 
Our model of an RTIL can be viewed as the complete association 
limit (CAL) of the $m$-component hard-sphere mixture  
with two sticky spots (patches) of the type $A$ and $B$, randomly placed on the surface of
each particle forming the cation chain. The model at hand will be recovered assuming infinitely
strong attraction between the patches of the type $B$ and $A$ located on the surface of
the particles of the type $i$ and $i+1$, respectively. Note that the size of each of the patches
is small enough to ensure that only one $A-B$ bond is formed. This feature of the model
allows us to present the multidensity OZ equation and AMSA closure relation in the following
form \cite{Protsykevytch,Kalyuzhnyi2001}:
\be
\hat{\bf h}_{ij}(k)=\hat{\bf c}_{ij}(k)+\rho\sum_l\hat{\bf c}_{il}(k){\balpha}
\hat{\bf h}_{lj}(k),
\label{OZ}
\ee
where $\hhat_{ij}(k)$, $\chat_{ij}(k)$ and $\balpha$ are the matrices,
\[
{\hat {\bf h}}_{ij}(k))=
\pmatrix{
{\hat h}^{00}_{ij}(k) & {\hat h}^{0A}_{ij}(k) &
{\hat h}^{0B}_{ij}(k) \cr
{\hat h}^{A0}_{ij}(k) & {\hat h}^{AA}_{ij}(k) &
{\hat h}^{AB}_{ij}(k) \cr
{\hat h}^{B0}_{ij}(k) & {\hat h}^{BA}_{ij}(k) &
{\hat h}^{BB}_{ij}(k) \cr},
{\hat {\bf c}}_{ij}(k)=
\pmatrix{
{\hat c}^{00}_{ij}(k) & {\hat c}^{0A}_{ij}(k) &
{\hat c}^{0B}_{ij}(k) \cr
{\hat c}^{A0}_{ij}(k) & {\hat c}^{AA}_{ij}(k) &
{\hat c}^{AB}_{ij}(k) \cr
{\hat c}^{B0}_{ij}(k) & {\hat c}^{BA}_{ij}(k) &
{\hat c}^{BB}_{ij}(k) \cr},
\]
\be
\hspace{4mm}
\balpha=
\pmatrix{
1 & 1 & 1 \cr
1 & 0 & 1 \cr
1 & 1 & 0 \cr},
\label{matrix}
\ee
with the
elements ${\hat h}^{\alpha \beta}_{ij}(k)$,
${\hat c}^{\alpha \beta}_{ij}(k)$ ($\alpha,\beta=0,A,B$)
being Fourier transforms of the partial
correlation functions $h^{\alpha \beta}_{ij}(r)$,
$c^{\alpha \beta}_{ij}(r)$, respectively. 
Note that $AA$ and $BB$ elements of the matrix $\balpha$ are equal 0, which  is the consequence of one bond per patch restriction, mentioned above.
Here
\be
{\bf c}_{ij}(r)=-{\bf E}\beta U_{ij}^{(C)}(r)+
{{\bf t}_{ij}\over 2\pi\sigma_{ij}}\delta(r-\sigma_{ij}),
\;\;\;r\le\sigma_{ij}
\label{cMSA}
\ee
\be
{\bf h}_{ij}(r)=-{\bf E},\;\;\;\;\;\;\;\;\;\;\;\;\;\;\;r<\sigma_{ij}
\label{hMSA}
\ee
where $E^{\alpha\beta}=\delta_{\alpha 0}\delta_{\beta 0}$,
$\beta=1/(k_BT)$, $k_B$ is Boltzmann constant,
\be
t_{ij}^{\alpha\beta}={1\over 2\rho}\left[\delta_{\alpha A}\delta_{\beta B}
{\delta_{i,j+1}\over\sigma_{i,i-1}}+\delta_{\alpha B}\delta_{\beta A}
{\delta_{i,j-1}\over\sigma_{i,i+1}}\right],\;\;\;i,j\ge 2,
\label{t1}
\ee
\be
t_{ij}^{\alpha\beta}=\left(\delta_{\alpha A}\delta_{\beta B}\delta_{i,j+1}+
\delta_{\alpha B}\delta_{\beta A}\delta_{i,j-1}\right)t,\;\;\;\;\;\;\;\;\;
i,j\le 2
\label{t2}
\ee
and
\be
\left.
t=2\pi\sigma_{12}^2x^2K^{(0)}_{ass}e^{G_{00}(\sigma_{12}^+)-\beta U^{(C)}(\sigma_{12}^+)}
g_{12}^{00}(\sigma_{12}^+)
\right\vert_{z_i=0}.
\label{Kass}
\ee
Note that in addition to the delta-function term in the closure relations
 for the correlation functions of the particles forming cation chain, eqs.
(\ref{cMSA}) and (\ref{t1}), the delta-function term
appears also in the correlation function for the anions and charged bead of the cations
${\bf c}_{12}(r)$, eqs. (\ref{cMSA}) and (\ref{Kass}).
This term is introduced to correct MSA closure for the effects of the ionic association
\cite{MF}. Here $\left.g^{00}_{ij}(\sigma_{12}^+)\right\vert_{z_i=0}$ 
is the contact value of the radial
distribution function $g^{00}_{ij}(r)=h^{00}_{ij}(r)+1$ at zero charges on the 
anion and cation bead, 
$\left.G_{00}(\sigma_{12}^+)=g^{00}_{12}(\sigma_{12}^+)-g^{00}_{12}(\sigma_{12}^+)\right\vert_{z_i=0}$, $K_{ass}^{(0)}$ is 
the association constant, $x$ is the fraction of free anions
(or cations) and in eq. (\ref{Kass}) we are using the exponential approximation \cite{MF}.
The upper indices $\alpha,\beta$ in the partial correlation functions 
${h}^{\alpha\beta}_{ij}(r),{c}^{\alpha\beta}_{ij}(r)$, which enter OZ equation 
(\ref{OZ}), are taking the values $0,A$ and $B$ and denote bonding states of the corresponding
particles \cite{Wertheim1986,Protsykevytch,Kalyuzhnyi2001}. The total partial correlation 
functions $h_{ij}^{\alpha\beta}(r)$ are related to the site-site total correlation functions 
$h_{ij}(r)$ by
\be
h_{ij}(r)=\sum_{\alpha\beta} h_{ij}^{\alpha\beta}(r)
\label{tot}
\ee
and the fraction of free anions(cations) $x$ follows from the solution of the mass action 
law type of equation
\be
\left.
4\pi\rho\sigma_{12}^2 x^2 K_{ass}^{(0)}e^{G_{00}(\sigma_{12}^+)-\beta U^{(C)}(\sigma_{12}^+)}g_{12}^{00}(\sigma_{12}^+)
\right\vert_{z_i=0}+x-1=0
\label{massaction}
\ee

The OZ equation (\ref{OZ}), AMSA closure relations (\ref{cMSA}) and (\ref{hMSA}) and equation
(\ref{massaction}) for $x$ form a close set of equations to be solved.

\section{Solution of the AMSA}

The solution of the AMSA for the models similar to that discussed above have been derived earlier  
\cite{Protsykevytch,Kalyuzhnyi2001} 
using Blum's version \cite{Blum1975,Bernard1995} of the Baxter factorization technique
\cite{Baxter}. We shall therefore omit here any details and present only the final expressions, suitable for the model in question. 
We will consider the general version of the model with different sizes of all monomers in the
system and two simplified versions, i.e. one with equal sizes of the neutral beads of the
cation and the other with only one neutral cation bead

According to 
\cite{Protsykevytch,Kalyuzhnyi2001,Kalyuzhnyi2007}, the solution of the set of equations
(\ref{OZ}), (\ref{cMSA}), (\ref{hMSA}) and (\ref{massaction}) can be reduced to the
solution of one nonlinear algebraic equation for Blum's screening parameter $\Gamma$
\be
\Gamma^2={\beta e^2\over 4
\epsilon\epsilon_0}
\rho\sum_{i=1}^m{\bf X}_i\balpha{\bf X}_i^T,
\label{Gamma}
\ee
where ${\bf X}_i=\left(X_i^0,X_i^A,X_i^B\right)$,
\be
X_i^0=\left[z_i-\eta^B\sigma_i^2\right]\Gamma_{\sigma_i},
\label{X0}
\ee
\be
X_i^\alpha=\sigma_i\left[\tau_i^\alpha(z)-\eta^B\tau_i^\alpha(\sigma^2)\right],
\;\;\;\;\;\;\alpha\neq 0\;\;(\alpha=A,B),
\label{XAB}
\ee
\be
\eta^B={{\pi\over 2\Delta}\rho\sum_{i=1}^m\sigma_i\left\{z_i\Gamma_{\sigma_i}+\sigma_i\left[
\tau_i^A(z)+\tau_i^B(z)\right]\right\}\over
1+{\pi\over 2\Delta}\rho\sum_{i=1}^m\sigma_i^2\left[\sigma_i\Gamma_{\sigma_i}
+\tau^A_i(\sigma^2)+\tau_i^B(\sigma^2)\right]},
\label{etaB}
\ee
\[
\tau_1^A(y)=0, 
\]
\[
\tau_2^A(y)=\rho\Gamma_{\sigma_1}\Gamma_{\sigma_2}y_1t,
\]
\[
\tau_i^A(y)={\Gamma_{\sigma_i}\Gamma_{\sigma_{i-1}}\over 2\sigma_{i,i-1}}
\left\{\sum_{l=3}^iy_{l-1}\left[
{\left(1-\delta_{li}\right)\over 2^{i-l}}\prod_{k=l}^{i-1}
{\sigma_k\Gamma_{\sigma_{k-1}} \over\sigma_{k,k-1}}+\delta_{li}\right]\;\;\;\;\;\;
\right.
\]
\[
\left.
+\rho\sigma_2y_1\Gamma_{\sigma_1}t\left[
{\left(1-\delta_{3i}\right)\over 2^{i-3}}
\prod_{k=3}^{i-1}{\sigma_k\Gamma_{\sigma_{k-1}}\over\sigma_{k,k-1}}+\delta_{3i}\right]\right\},
\;\;\;\;3\le i\le m,
\]
\[
\tau_1^B(y)=\rho\Gamma_{\sigma_1}\Gamma_{\sigma_2}t\sum_{l=2}^my_l
\left[
{\left(1-\delta_{l2}\right)\over 2^{l-2}}
\prod_{k=2}^{l-1}{\sigma_k\Gamma_{\sigma_{k+1}}\over\sigma_{k,k+1}}
+\delta_{l2}\right]
\]
\[
\tau_i^B(y)={\Gamma_{\sigma_i}\Gamma_{\sigma_{i+1}}\over 2\sigma_{i,i+1}}
\sum_{l=i+1}^my_{l}\left[
{\left(1-\delta_{l,i+1}\right)\over 2^{l-i-1}}
\prod_{k=i+1}^{l-1}
{\sigma_k\Gamma_{\sigma_{k+1}}\over\sigma_{k,k+1}}+\delta_{l,i+1}\right],
\;\;\;\;2\le i<m,
\]
\[
\tau^B_m(y)=0.
\]

Here $\beta^*=\beta e^2/(4\pi\epsilon\epsilon_0)$,
$\Gamma_{\sigma_i}=\left(1+\sigma_i\Gamma\right)^{-1}$ and $y$ is taking the values
either $z$ or $\sigma^2$.

\subsection{Model with equal sizes of the cation neutral beads (model $C$)}

Substantial simplification of the above expressions for $\tau_i^\alpha(y)$ occurs
for the model with equal hard-sphere sizes of all $m-2$ neutral beads of the cation,
i.e. $\sigma_i=\sigma_n$ for $i\geq 3$. We have:
\[
\tau^A_1(y)=0,
\]
\[
\tau_2^A(y)=\rho\Gamma_{\sigma_a}\Gamma_{\sigma_c}y_at
\]
\[
\tau_3^A(y)={\Gamma_{\sigma_n}\Gamma_{\sigma_c}\over 2\sigma_{nc}}\left(
y_c+\rho\sigma_cy_a\Gamma_{\sigma_a}t\right)
\]
\[
\tau_i^A(y)={\Gamma_{\sigma_n}^2\over 2\sigma_{n}}\left[\left(
y_c+\rho\sigma_cy_a\Gamma_{\sigma_a}t\right){\sigma_n\Gamma_{\sigma_c}\over 2\sigma_{nc}}
\left({\Gamma_{\sigma_n}\over 2}\right)^{i-4}+y_n\sum_{l=4}^i
\left({\Gamma_{\sigma_n}\over 2}\right)^{i-l}\right],
\;4\leq i\leq m,
\]
\[
\tau_1^B(y)=\rho\Gamma_{\sigma_a}\Gamma_{\sigma_c}t\left[y_c+y_n{\sigma_c\over\sigma_{cn}}
\sum_{l=3}^m\left({\Gamma_{\sigma_n}\over 2}\right)^{l-2}\right],
\]
\[
\tau_2^B(y)={\Gamma_{\sigma_c}\Gamma_{\sigma_n}\over 2\sigma_{cn}}y_n
\sum_{l=3}^m\left({\Gamma_{\sigma_n}\over 2}\right)^{l-3},
\]
\[
\tau_i^B(y)={\Gamma_{\sigma_n}^2\over 2\sigma_n}y_n
\sum_{l=i+1}^m\left({\Gamma_{\sigma_n}\over 2}\right)^{l-i-1},
\;\;\;\;\;3\leq i<m,
\]
\[
\tau_m^B(y)=0.
\]

Note that in the above expressions for the sake of convenience we are using a slightly modified
notation, i.e $\sigma_1=\sigma_a$, $\sigma_2=\sigma_c$, $y_1=y_a$, $y_2=y_c$ and 
$y_i=y_n\;(i>2)$.

\subsection{Model with cation having one neutral bead (model $D$)}

Further simplification is possible for the model with the cation represented by the dimer, i.e.

\[
\tau_1^A(y)=0,\;\;\;\;\tau_2^A(y)=\rho\Gamma_{\sigma_a}\Gamma_{\sigma_c}y_at,
\;\;\;
\tau_3^A(y)={\Gamma_{\sigma_c}\Gamma_{\sigma_n}\over 2\sigma_{nc}}
\left(y_c+\rho\sigma_cy_a\Gamma_{\sigma_a}t\right),
\]
\[
\tau_1^B(y)=\rho\Gamma_{\sigma_a}\Gamma_{\sigma_c}t\left(y_c+y_n
{\sigma_c\over 2\sigma_{nc}}\Gamma_{\sigma_n}\right),\;\;\;
\tau_2^B(y)={\Gamma_{\sigma_c}\Gamma_{\sigma_n}\over 2\sigma_{nc}}y_n,\;\;\;
\tau_3^B(y)=0.
\]

Using these expressions for $\tau_i^\alpha$ in expressions (\ref{X0}), (\ref{XAB}) 
and (\ref{etaB}) for $\eta^B$ and ${\bf X}_i^\alpha$, respectively, we have: 
\be
\eta^B={{\pi\rho\over 2\Delta} z\left\{\sigma_c\Gamma_{\sigma_c}-\sigma_a\Gamma_{\sigma_a}+
\left[
{\sigma_n^2\over 2\sigma_{nc}}\Gamma_{\sigma_n}+
\rho\Gamma_{\sigma_a}t\left(\sigma_a^2-\sigma_c^2-{\sigma_c\sigma_n^2\over 2\sigma_{nc}}\Gamma_{\sigma_n}\right)\right]\Gamma_{\sigma_c}\right\}\over 
1+{\pi\rho\over 2\Delta}\left\{\sum_{q=a}^n\sigma_q^3\Gamma_{\sigma_q}
+\sigma_c^2\left[{\sigma_n^2\over  \sigma_{nc}}\Gamma_{\sigma_n}+
\sigma_a^2\rho\Gamma_{\sigma_a}t
\left(2+{\sigma_n^2\over \sigma_{cn}\sigma_c}\Gamma_{\sigma_n}\right)
\right]\Gamma_{\sigma_c}\right\}},
\label{etaB3}
\ee
where $q$ is taking the values $a,c,n$ and
\be
{\bf X}_1={\bf X}_a=\Gamma_{\sigma_a}\left\{-z-\eta^B\sigma_a^2,\;0,\;
\sigma_a\Gamma_{\sigma_c}t\left[z-\sigma_c\eta^b\left(\sigma_c+
{\sigma_n^2\over 2\sigma_{cn}}\Gamma_{\sigma_n}\right)\right]\right\},
\label{X1}
\ee
\be
{\bf X}_2={\bf X}_c=\Gamma_{\sigma_c}
\left\{z-\eta^B\sigma_c^2,\;\;\;-\sigma_c\rho\Gamma_{\sigma_a}t\left(z
+\eta^B\sigma_a^2\right),\;\;\;
-{\sigma_c\sigma_n^2\over 2\sigma_{an}}\eta^B\Gamma_{\sigma_n}\right\},
\label{X2}
\ee
\be
{\bf X}_3={\bf X}_n=\Gamma_{\sigma_n}
\left\{-\eta^B\sigma_n^2,\;{\sigma_n\over 2\sigma_{nc}}
\left[z-\eta^B\sigma_c^2-\rho\sigma_c\Gamma_{\sigma_a}t
\left(z+\eta^B\sigma_a^2\right)\right],\;0\right\}.
\label{Xn}
\ee

\section{Structural and thermodynamic properties}

The equation for the fraction of free anions $x$ (eq. (\ref{massaction}))
 includes the contact values
of the radial distribution function $g_{12}^{00}(\sigma_{12}^+)$ of the original version
of the model and the version of the model with $z_i=0$, i.e 
$\left. g_{12}^{00}(\sigma_{12}^+)\right\vert_{z_i=0}$. We have 
\cite{Protsykevytch,Kalyuzhnyi2001}:
\be
\sigma_{ij}g_{ij}^{00}(\sigma_{ij}^+)=\left. 
\sigma_{ij}g_{ij}^{00}(\sigma_{ij}^+)\right\vert_{z_i=0}-{\beta e^2\over 4\pi
\epsilon\epsilon_0}X_i^0X_j^0.
\label{cont}
\ee
In the framework of the present AMSA closure the contact value of the radial distribution 
function of the model at zero charges coincide with the Percus-Yevick contact value of the
corresponding radial distribution function of the $m$-component mixture of hard spheres
$g_{ij}^{(hs)}$, i.e.
\be
\left. g_{ij}^{00}(\sigma_{ij}^+)\right\vert_{z_i=0}=g_{ij}^{(hs)}(\sigma_{12}^+)=
{1\over 1-\eta}+{\pi\sigma_i\sigma_j\over 4\sigma_{ij}}{\rho
\sum_{l=1}^m\sigma_l^2\over \left(1-\eta\right)^2},
\label{contPY}
\ee
where $\eta=\pi\rho/6\sum_{l=1}^m\sigma_l^3$.

Following 
\cite{Protsykevytch,Kalyuzhnyi2001} for the excess internal energy 
of the model $\Delta E$ we have
\be
\beta{\Delta E\over V}={\beta e^2\over 4\pi
\epsilon\epsilon_0}\rho z\left[{1\over\sigma_c}\left(\sum_{\alpha=0}^BX_c^\alpha-z\right)-{1\over\sigma_a}\left(\sum_{\alpha=0}^BX_a^\alpha+z\right)\right].
\label{energy}
\ee
The Helmholtz free energy of the model $A$ can be written as a sum of four terms, i.e. the
ideal gas contribution, a contribution due to bonding or the so-called
mass action law contribution \cite{Bernard,Jiang2002} $\Delta A^{(MAL)}$ and contributions 
due to the hard sphere and electrostatic
interactions, $\Delta A_{hs}$ and $\Delta A_{el}$, respectively:
\be
{\beta A\over V}={\beta A^{(id)}\over V}+{\beta\Delta A^{(MAL)}\over V}+
{\beta\Delta A^{(hs)}\over V}+{\beta\Delta A^{(el)}\over V},
\label{A_gen}
\ee
where
\be
\beta A^{(id)}/V=2\rho\left(\ln{\rho}-1\right)
\label{ideal}
\ee,
\be
{\beta\Delta A^{(MAL)}\over V}=\rho\left(\ln{x}-{1\over 2}x+{1\over 2}\right)
-\rho\sum_{i=2}^{m-1}\ln{\left[g_{i,i+1}^{(hs)}(\sigma_{i,i+1}^+)\right]}.
\label{AMAL}
\ee
For the hard-sphere contribution we are using the Boublik-Mansoori-Carnahan-Starling-Leland (BMCSL) 
expression \cite{Boublik1970,Mansoori1971}, and electrostatic contribution can be calculated
numerically, using the coupling constant integration. 

All the rest of thermodynamic properties can be obtained using the standard thermodynamic relations.
In particular for the pressure $P$ and for the anion and cation chemical potentials $\mu_a$ and 
$\mu_c$ we have:
\be
P=-{\partial A\over\partial V},\;\;\;\;
\rho\left(\mu_a+\mu_c\right)={\beta A\over V}+
\beta P.
\label{rest}
\ee

\section{Calculation of the phase diagram}

The liquid-gas phase diagram was calculated extending the method proposed earlier 
\cite{Kalyuzhnyi1998,Kalyuzhnyi2000,Jiang2002}. 
We assume that at the temperatures close to the phase
transition all anions and cations are dimerized, so that the system properties can be
described using CAL \cite{Kalyuzhnyi1998}, i.e. 
$K_{ass}^{(0)}\rightarrow\infty$ and $t=1/(2\rho\sigma_{ac})$. 
This assumption is based on the MC computer simulation observations, which
suggest that in the coexisting phases the fraction of nonbonded ions is negligible and
the phase diagram of the RPM can be reasonably well represented
by the phase diagram of the corresponding fluid of dimers formed by the oppositely charged
ions \cite{Shelley1995,Camp1999,Daub2003,Romero2000}. 
In addition we follow Blum et al. \cite{Jiang2002} 
and in a spirit of Wertheim's multidensity thermodynamic perturbation theory 
\cite{Wertheim1984,Wertheim1986}
assume different approximations for different terms in the expression for 
Helmholtz free energy (\ref{A_gen}), i.e. we calculate $\Delta A^{(el)}$ 
using the $\Gamma$ parameter obtained in the complete dissociation limit (CDL) 
($K^{(0)}_{ass}=0$ and $t=0$). According to Blum et al. \cite{Jiang2002}
this combination of the CAL approximation for $\Delta A^{(MAL)}$ and CDL approximation for 
$\Delta A^{(el)}$ can be seen as an $ad\;hoc$ 
interpolation between the AMSA and the simple interpolation scheme  of Stell and Zhou
\cite{Stell1989}, which gives the most accurate prediction for the phase behavior
of the PM of electrolytes \cite{Jiang2002,Qin2004}. Note that in the CDL $\Gamma$ contains contribution due to the presence of the neutral
beads of the cation chain and cannot be reduced to regular MSA $\Gamma$ parameter.
Taking into account these two assumptions 
we have:
\[
\left.
{\beta\Delta A^{(MAL)}\over V}\right\vert_{K^{(0)}_{ass}\rightarrow\infty}=
\]
\be
-\rho\left\{\ln{\rho}-1
+\sum_{i=1}^{m-1}\ln{\left[g_{i,i+1}^{(hs)}(\sigma_{i,i+1}^+)\right]}
+\beta U_{12}^{(C)}(\sigma_{12})+G_{00}^{(\infty)}(\sigma_{12}^+)\right\}
\label{AMAL_inf}
\ee
where
$G_{00}^{(\infty)}(\sigma_{12}^+)=G_{00}(\sigma_{12}^+)|_{K_{ass}^{(0)}\rightarrow\infty}$, and 
\be
\left.
{\beta\Delta A^{(el)}\over V}\right\vert_{K^{(0)}_{ass}=0}
={\beta\Delta E^{(0)}\over V}+{\left(\Gamma^{(0)}\right)^3\over 3\pi},
\label{freeG}
\ee
where
$\left. \Gamma^{(0)}=\Gamma\right\vert_{K^{(0)}_{ass}=0}$ and $E^{(0)}=E|_{K^{(0)}_{ass}=0}$.
The corresponding expression for the pressure $P$ is:
\be
\beta P=\rho+\beta\Delta P^{(MAL)}+\beta\Delta P^{(hs)}+\beta\Delta P^{(el)},
\label{PP}
\ee
where
\be
\beta\Delta P^{(MAL)}=-\rho^2\left[\sum_{i=1}^{m-1}
{\partial\ln{\left[g_{i,i+1}^{(hs)}(\sigma_{i,i+1}^+)\right]}\over\partial\rho}
+{\partial G_{00}^{(\infty)}(\sigma_{12}^+)\over\partial\rho}\right].
\label{PMAL}
\ee
For the hard-sphere contribution $\Delta P^{(hs)}$ we are using 
BMCSL expression \cite{Boublik1970,Mansoori1971}, and electrostatic contribution
$\Delta P^{(el)}$ is:
\be
\beta\Delta P^{(el)}=-{\left(\Gamma^{(0)}\right)^3\over 3\pi}-{\beta e^2\over 2\pi^2
\epsilon\epsilon_0}
\left(\eta^B_0\right)^2,
\label{Pel}
\ee
where
$\eta^B_0=\eta^B|_{K_{ass}^{(0)}=0}$. 

As usual the phase diagram was calculated from the equality of the pressure and chemical potentials
in the coexisting phases.

\section{Monte Carlo simulations}

Monte Carlo computer simulations were performed in the canonical ensemble using Molsim
software \cite{molsim}. A model system had 100 particles of each type. Electrostatic interactions 
were calculated using the Ewald technique. During equilibration $4\cdot10^6$ attempted moves per particle 
were performed and followed by $40-120\cdot10^6$ attempts for production run. Cations were both displaced and 
rotated simultaneously during a trial move. 
The excess chemical potential was calculated via Widom's test particle insertion method. An electroneutral pair 
(one anion and one cation) was randomly inserted after every 20 trial moves per particle. Inserted test cations 
were oriented randomly. At each temperature simulations were carried out at a total of 20 densities. Ideal, hard-sphere, and 
excess contribution were added together to obtain the total chemical potential. Coexisting densities were determined using the Maxwell 
construction in the $\mu - \rho$ plane. 
The Widom method is known to perform best at low to moderate densities. We were able to insert linear cations with 
$m_c=m-1=2$ and $m_c=3$ at all densities studied. However, a fraction of successful test particle insertions dropped significantly 
for larger cations ($m_c>3$ or $\sigma_n>\sigma_a$) at larger densities, preventing us from obtaining phase diagrams for these models.

\section{Results and discussion}

We have studied the phase behavior of two versions of the primitive models of RTIL proposed.
The first version (model $D$) is represented by the model with cations modeled by  
dimers with the neutral bead of different sizes ($\sigma_a\leq\sigma_n\leq 3\sigma_a$) and the 
second one (model $C$) is a the model with cations represented by the flexible chains of $m_c$ 
tangentially bonded hard- sphere monomers ($1\leq m_c\leq 8$) of the same size. In all cases 
studied the sizes of the anions and charged beads of the cations were chosen to be equal, 
i.e. $\sigma_a=\sigma_c=\sigma$. In what follows the model parameters are expressed in 
terms of the dimensionless quantities: reduced density $\rho^*=\rho\sigma_a^3$ and reduced 
temperature $T^*=4\pi\epsilon\epsilon_0 k_BT/e^2$.

\begin{figure*} 
\begin{center}
\includegraphics[width = 0.9\textwidth]{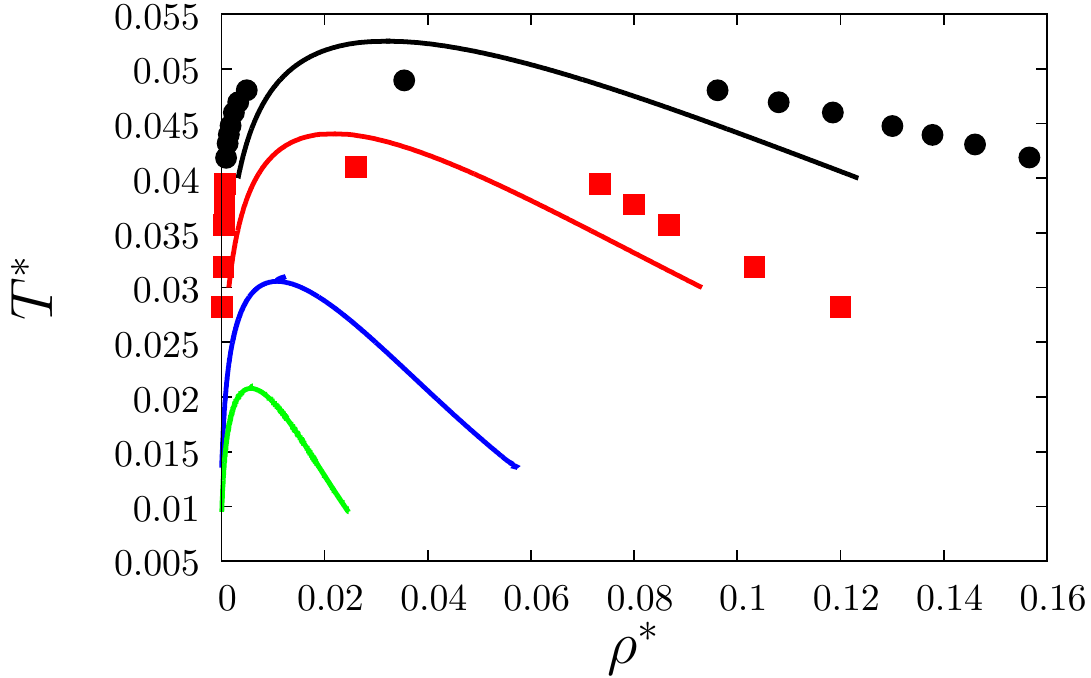}
\caption{Liquid-gas phase diagram of model $D$, with the cation represented by the dimer
($m_c=2$) with different hard-sphere sizes of the neutral bead, i.e. 
$\sigma_n=0,\sigma,2\sigma,3\sigma$ (from the top to the bottom) shown by the solid black, red, blue and green lines, respectively. 
Here $\sigma_c=\sigma_a=\sigma$.
Black circles and red squares represent computer simulation results for the models with $\sigma_n=0$ (RPM \cite{panagiotop2002}) and with $\sigma_n=\sigma$, respectively.}
\label{fig2}
\end{center}
\end{figure*}

\begin{figure*} 
\begin{center}
\includegraphics[width = 0.9\textwidth]{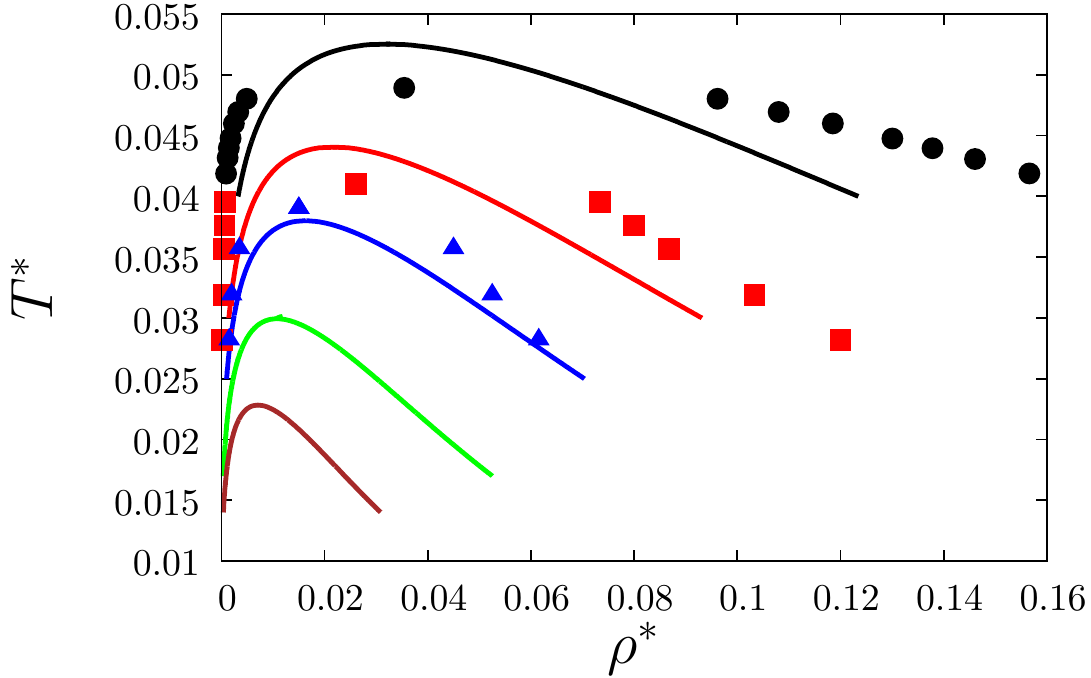}
\caption{Liquid-gas phase diagram of model $C$, with equal hard-sphere sizes for all
monomeric units ($\sigma_a=\sigma_c=\sigma_n=\sigma$) and different lengths of the cation chains
$m_c=1,2,3,5,8$ (from the top to the bottom) shown by the solid black, red, blue, green and
brown lines, respectively. 
Black circles, red squares and blue triangles represent computer 
simulation results for the models with $m_c=1$ (RPM, \cite{panagiotop2002}), $m_c=2$ and
$m_c=3$, respectively.}
\label{fig3}
\end{center}
\end{figure*}

In figure \ref{fig2} we present our theoretical results for the model $D$ with 
$\sigma_n=\sigma,2\sigma,3\sigma$ and in figure \ref{fig3} for model $C$ with $m_c=2,3,5,8$.
These results are compared against computer simulation results for the models with
$\sigma_n=\sigma$
 (model $D$) and with $m_c=2,3$ (model $C$). In addition, we also
show theoretical and computer simulation results for the phase diagram of electrolyte RPM 
($\sigma_n=0$, $m_c=1$). In general theoretical predictions for the RTIL models and for RPM are of the 
same order of accuracy. The theory gives relatively accurate results for the 
critical density, predictions for the critical temperature are less accurate.  
For the RPM, the theory to be around 7\% too high the critical temperature and with
the increase of the neutral bead size $\sigma_n$ (figure \ref{fig2}) or cation chain length $m_c$ 
(figure \ref{fig3}) this disagreement gradually increases.

\begin{figure*} 
\begin{center}
\includegraphics[width = 0.9\textwidth]{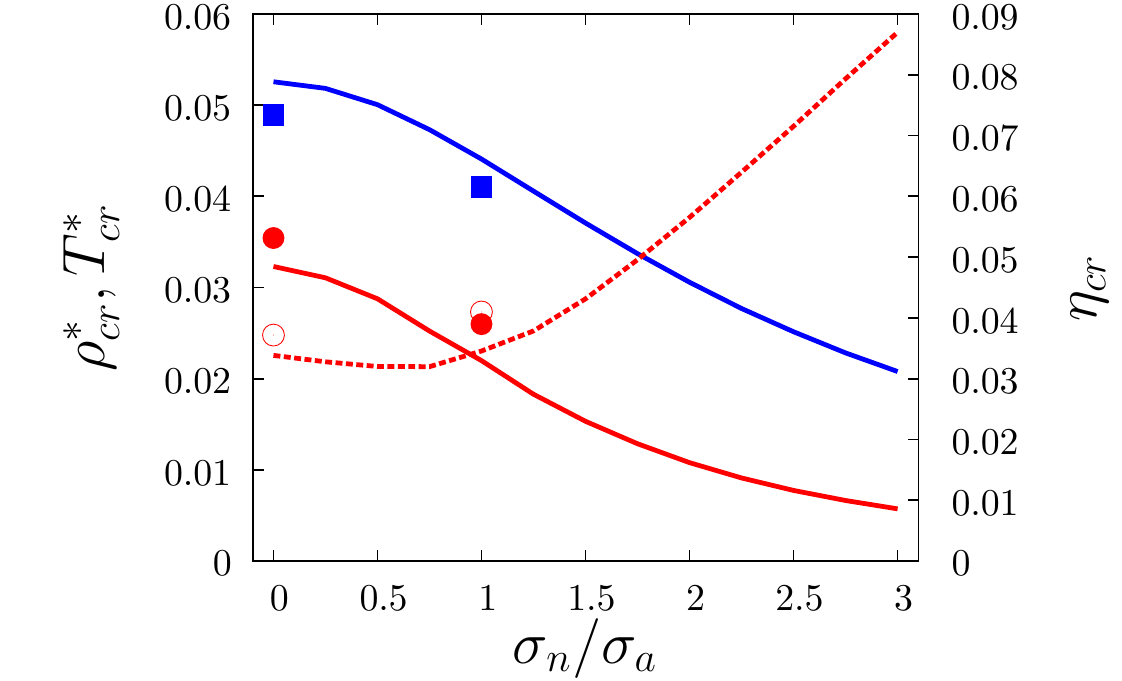}
\caption{Critical density $\rho_{cr}^*$ (red solid line), critical packing fraction 
$\eta_{cr}$ (red dashed line) and critical temperature $T^*$ (blue solid line) as a
functions of the size of the cation neutral bead $\sigma_n$.
Red filled circles, blue filled squares and red open circles represent computer 
simulation results for the critical density, critical temperature and critical packing fraction, respectively. Computer simulation results for the model with $\sigma_n=0$ (RPM)
are taken from Ref. \cite{panagiotop2002}.}
\label{fig4}
\end{center}
\end{figure*}

\begin{figure*} 
\begin{center}
\includegraphics[width = 0.9\textwidth]{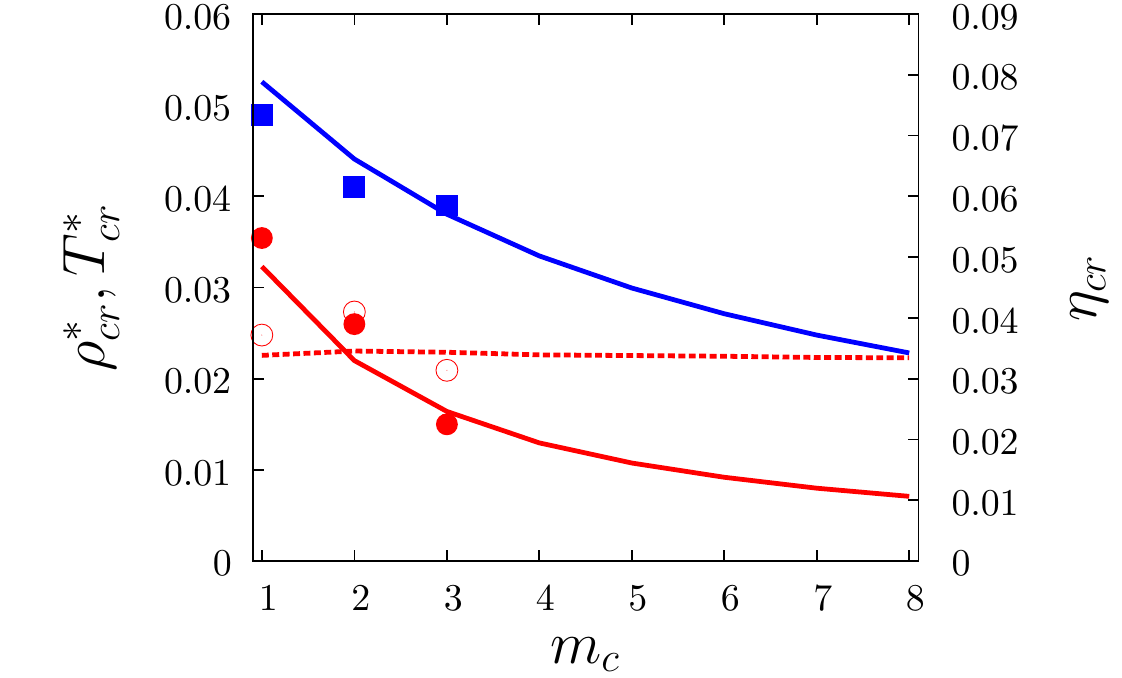}
\caption{Critical density $\rho_{cr}^*$ (red solid line), critical packing fraction 
$\eta_{cr}$ (red dashed line) and critical temperature $T^*$ (blue solid line) as a 
functions of the cation chain length $m_c=m-1$.
Red filled circles, blue filled squares and red open circles represent computer 
simulation results for the critical density, critical temperature and critical packing fraction, respectively. Computer simulation results for the model with $m_c=1$ (RPM)
are taken from Ref. \cite{panagiotop2002}.
}
\label{fig5}
\end{center}
\end{figure*}

With the increase of the model asymmetry due to the increase of $\sigma_n$ or
$m_c$  the phase envelope and critical point are moving towards lower temperatures and lower densities. This shift of the phase diagram is reflected in figures \ref{fig4} and 
\ref{fig5}, where we show the critical density and critical temperature as a function of the 
neutral bead 
size $\sigma_n$ for the model $D$ and as a function of the cation chain length $m_c$ for the 
model $C$, respectively. 
Similar behavior was observed for the primitive model of RTIL with cations represented
by spherocylinders \cite{Martin2009}, i.e. with the increase of spherocylinder length critical temperature
and critical density decrease. According to Martin-Betancourt et al. \cite{Martin2009}
this effect is of entropic origin, since the presence of uncharged tails reduces the
number of energetically favorable configurations of the ions. 
Increasing $\sigma_n$ and/or $m_c$ reduces the number of
configurations in which when anion and charged bead of the cation are in contact. 
This feature of the model reduces its ability to form clusters and as a result both
critical temperature and critical density decrease. Increase of the critical temperature, 
caused by the increase of dispersion attraction due to chain length increase
is a common feature of nonionic fluids such as alkanes or alcohols \cite{27}.
This difference in the behavior of the critical temperature is a clear indication that Coulomb
interaction is the major driving force of the phase transitions in RTIL 
and to a first
approximation dispersion forces can be neglected (as it is done in primitive models of RTIL).
In the same figures we present also the critical packing fraction
$\eta_{cr}$ $vs$ $\sigma_n$ (figure \ref{fig4}) 
and $\eta_{cr}$ $vs$ $m_c$ (figure \ref{fig5}).
For model $C$, $\eta_{cr}$ appears to be almost independent on
the cation chain length (including RPM), i.e. computer simulation and theory give 
$\eta_{cr}\approx 0.04$ and $\eta_{cr}\approx 0.034$, respectively. The same value 
of the critical packing fraction $\eta_{cr}\approx 0.04$ was obtained for the primitive 
models of the RTIL studied earlier using computer simulation methods 
\cite{Martin2009,Camp2011}. For model
$D$, the situation is different, i.e. here with the increase of the neutral bead size
$\sigma_n$ for $\sigma_n/\sigma_a\leq 0.7$  $\eta_{cr}$ slightly decreases and for 
$0.7 < \sigma_n/\sigma_a\leq  3$ it shows substantial increase.

\begin{figure*} 
\begin{center}
\includegraphics[width = 0.9\textwidth]{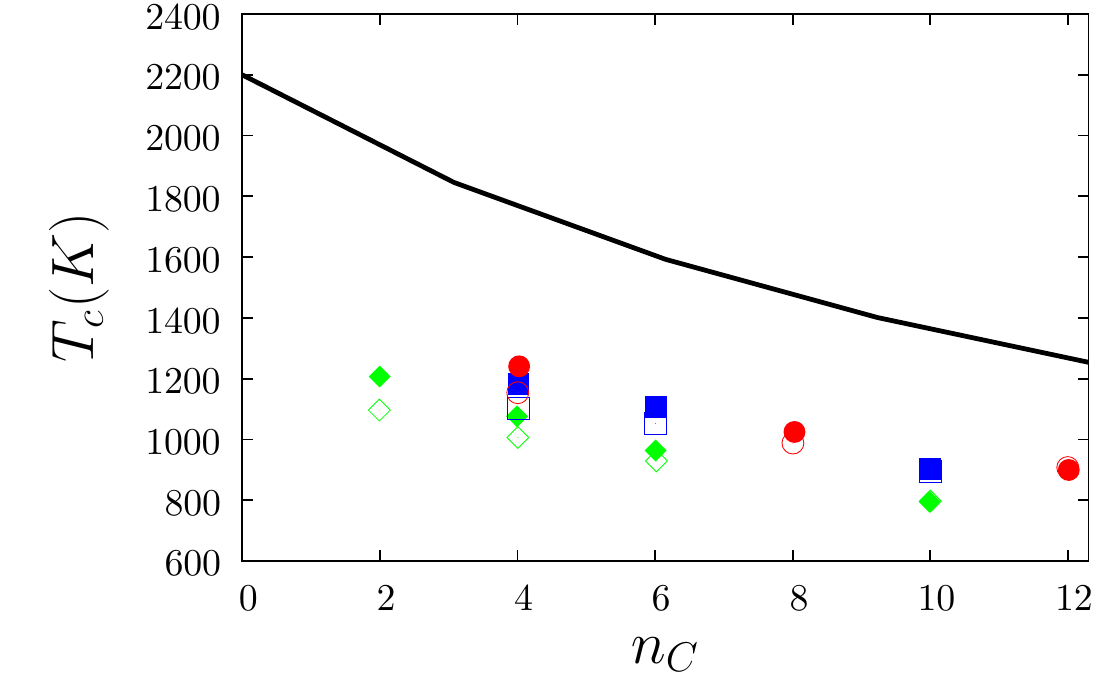}
\caption{Critical temperature $T_{cr}(K)$ as a function of the number of carbons in the imidazolium-based ionic liquids. Symbols denote experimental results
\cite{Rebelo2005} for
[C$_{\rm n}$mim][BF$_4$] (red circles), [C$_{\rm n}$mim][PF$_6$] (blue squares) and 
[C$_{\rm n}$mim][Ntf$_2$] (green diamonds) and solid line denotes results of our theory. 
Here filled symbols represent results obtained using the E{\"o}tvos relation and open symbols denote results calculated using the Guggenheim relation.}
\label{fig6}
\end{center}
\end{figure*}

Finally in figure \ref{fig6} 
we compare our results for the critical temperature of the model 
$C$ against corresponding results obtained by extrapolating the experimental data for the 
surface tension and density of a homologous series of imidazolium-based ionic liquids, i.e.
[C$_{\rm n}$mim][BF$_4$], [C$_{\rm n}$mim][PF$_6$] and [C$_{\rm n}$mim][Ntf$_2$]
\cite{Rebelo2005}. Since these ionic liquids are not stable at higher temperatures
Rebelo et al. \cite{Rebelo2005} have used the Guggenheim and E{\"o}tvos empirical relations
\cite{Shereshevsky1935,Guggenhein1945} to calculate the critical temperature. 
Our goal here is to verify the ability of the model and theory to give at least 
qualitatively correct description of the experimentally observed behavior.
We have not made any attempts to optimize the model parameters and for the sake of simplicity 
follow Martin-Betancourt et al. \cite{Martin2009} assuming the following values: 
$\sigma=4\;\AA$ and $L=1.3n_C(\AA)$, where $L$ is the chain length and $n_C$ is the number
of carbons. Here $L$ is the distance 
between the centers of the terminal beads for completely stretched chain expressed in Angstroms. Note that for our model $L$ can take only values that are multiples of 
$\sigma$. The critical temperature for the intermediate values of $L$ was calculated
via linear interpolation between the values obtained for the chain length $L$ being 
a multiple of $\sigma$. Although our results are about 60\% higher than experimental
results, the model and the theory proposed are able to give qualitatively correct behavior,
i.e. with the increase of $n_C$ both the theoretical and computer simulation critical temperature decreases. For the longer chains this decrease becomes slightly less steep. Further improvement of our 
results can be achieved by recognizing that the dielectric permittivity 
$\epsilon$
in the
expression for the Coulomb potential (\ref{C}) can take different values. In the current study
we assume that 
$\epsilon=1$.
For imidazolium-based RTIL with ions lacking polar groups the dielectric
permittivity is related to refractive index $n_D$, i.e. 
$\epsilon=n^2_D$.   
In the recent paper, Lu et al. \cite{Forsman2016} assume that 
$\epsilon=2$. 
Our theory appears to be in a very good agreement with experiment assuming
that 
$\epsilon \approx 1.60-1.67$ 
(see figure \ref{fig6}).

\section{Conclusions}

In this paper we proposed several versions of primitive models for RTILs and developed
a theory for their description using the analytical solution of the AMSA.
Solution of the corresponding multidensity Ornstein-Zernike equation, supplemented by the 
AMSA, reduces to the solution of only one nonlinear algebraic equation for Blum's screening
parameter $\Gamma$. The theory is used to study the liquid-gas phase behavior of 
two versions of the model, i.e. models with cations represented by dimers with a
neutral bead of different size and by chains with the neutral beads of the 
same size, respectively. We generated a set of computer simulation results for the liquid-gas 
phase diagrams of the models. Theoretical predictions for these phase diagrams appear to
be in reasonably good agreement with computer simulation predictions. 
It is demonstrated that the models and theory are able to reproduce experimentally observed
trends in the phase behavior of RTILs, in particular, the decrease of the critical temperature and 
critical density with the increase of the asymmetry of the model in its shape and position 
of the charge.

\section{Acknowledgment}

J.R. thanks the Slovenian Research Agency for its financial support through grant P1-0201 and
M.H. acknowledges support from the European Union's Horizon 2020 research and innovation pogramme
under the Marie Sklodowska-Curie (grant No.734276) and the State Fund for Fundamental Research (project NF73/3/26-2017).

\newpage

\newpage

\newpage

\clearpage

\newpage

\clearpage

\section*{References}

\begin{thebibliography}{99}

\bibitem{Waisman1972a} E. Waisman, J.L. Lebowitz, J.Chem.Phys. 56 (1972) 3086.
\bibitem{Waisman1972b} E. Waisman, J.L. Lebowitz, J.Chem.Phys. 56 (1972) 3093.
\bibitem{Blum1975} L. Blum, Mol. Phys. 30 (1975) 1529.
\bibitem{Blum1977} L. Blum, J.S. H{\o}ye, J. Phys. Chem. 81 (1977) 1311.
\bibitem{real_review} E.J. Maginn, J. Phys.: Condens. Matter 21 (2009) 373101.
\bibitem{Malvaldi2007} M. Malvaldi, C. Chiappe, J. Phys.: Condens. Matter 20 (2008) 035108. 
\bibitem{Spohr2009} H.V. Spohr, G.N. Patey, J.Chem.Phys. 130 (2009) 104506.
\bibitem{Martin2009} M. Martin-Betancourt, J.M. Romero-Enrique, L.F. Rull,
J.Phys.Chem. B, 113 (2009) 9046.
\bibitem{Camp2011} G.C. Ganzenmüller, P.J. Camp, Condens.Matter Phys. 14 (2011) 33602.
\bibitem{Lindergerg2014} E.K. Lindenberg, G.N. Patey, J.Chem.Phys. 140 (2014) 104504.
\bibitem{Linderberg2015} E.K. Lindenberg, G.N. Patey, J.Chem.Phys. 143 (2015) 024508.
\bibitem{Alcantara2016} W. Silvestre-Alcantara, L. Bhuiyan, S. Lamperski, M. Kaja, 
D. Henderson, Condens.Matter Phys. 19 (2016) 13603.
\bibitem{Forsman2016} H. Lu, B. Li, S. Nordholm, C.E. Woodward, J. Forsman,
J.Chem. Phys. 145 (2016) 234510.
\bibitem{Wertheim1984} M.S. Wertheim, J.Stat.Phys. 35 (1984) 19,35.
\bibitem{Wertheim1986} M.S. Wertheim, J.Stat.Phys. 42 (1986) 459,477.
\bibitem{amsa} M.F. Holovko, Y.V. Kalyuzhnyi, Mol.Phys. 73 (1991) 1145. 
\bibitem{Protsykevytch} I.A. Protsykevytch, Y.V. Kalyuzhnyi, M.F. Holovko, L. Blum,
J.Molec.Liq. 73,74 (1997) 1.
\bibitem{Kalyuzhnyi2001} Y.V. Kalyuzhnyi, P.T. Cummings, J. Chem. Phys. 115 (2001) 540.
\bibitem{Stillinger1968} F. Stillinger, R. Lovett, J.Chem.Phys. 48 (1968) 3858.
\bibitem{Ebeling1971} W. Ebeling, M. Grigo, Ann.Phys.(Leipzig) 37 (1980) 21.
\bibitem{Stell1976} G. Stell, K.C. Wu, B. Larsen, Phys.Rev.Lett. 37 (1976) 1369.
\bibitem{Vorontsov1970} N.P. Vorontsov-Veliaminov, A.M. El’yashevich, L.A. Morgenshtern,
V.P. Chasovskikh, Teplofiz.Vys.Temp. 8 (1970) 277.
\bibitem{Stell1995} G. Stell, J. Stat. Phys. 78 (1995) 197.
\bibitem{Weingartner2001} H. Weingartner, W. Schroer, Adv. Chem. Phys. 116 (2001) 1.
\bibitem{Hynninen2008} A.P. Hynninen, A.Z. Panagiotopoulos, Mol. Phys. 106 (2008) 2039.
\bibitem{caillol} J.M. Caillol, D. Levesque, J.-J. Weis, J.Chem.Phys. 116 (2002) 10794.
\bibitem{panagiotop2002} A.Z. Panagiotopoulos, J.Chem.Phys. 116 (2002) 3007.
\bibitem{panagiotop2003} D.W. Cheong, A.Z. Panagiotopoulos, J.Chem.Phys. 119 (2003) 8526.
\bibitem{Fisher1993} M.E. Fisher, Y. Levin, Phys. Rev. Lett. 71 (1993) 3826.
\bibitem{Fisher1994} M.E. Fisher, J. Stat, Phys. 75 (1994) 1.
\bibitem{Levin1996} Y. Levin, M.E. Fisher, Physica A 225 (1996) 164.
\bibitem{Patsahan2006} O. Patsahan, I. Mryglod, T. Patsahan, J. Phys.-Cond. Matt. 18 (2006) 10223.
\bibitem{Patsahan2010a} O.V. Patsahan, T.M. Patsahan, Cond. Matt. Phys. 13 (2010) 23004. 
\bibitem{Patsahan2010b} O.V. Patsahan, T.M. Patsahan, Phys. Rev. E 81 (2010) 031110. 
\bibitem{Jiang2002} J. Jiang, L. Blum, O. Bernard, J.M. Prausnitz, S.I. Sandler, 
J.Chem.Phys. 116 (2002) 7977.
\bibitem{Holovko2017a} M. Holovko, T. Patsahan, O. Patsahan, J.Molec.Liq. 228 (2017) 215.
\bibitem{Holovko2017b} M.F. Holovko, T.M. Patsahan, O.V. Patsahan, 
J.Molec.Liq. 235 (2017) 53.
\bibitem{Qin2004} Y.~Qin, J.~M. Prausnitz, J.~Chem.~Phys. 121 (2004) 3181.
\bibitem{MF} M.F. Holovko In "Ionic Soft Matter:applications", D.Henderson, M.Holovko, 
A.Trokhymchuk (Eds), Springer, Dordrecht, Netherland, 2005, vol.206, pp 45-81.
\bibitem{Bernard1995} L. Blum, O. Bernard, J.Stat.Phys. 79 (1995) 569.
\bibitem{Baxter} R.J. Baxter, J. Chem. Phys. 49 (1968) 2770.
\bibitem{Kalyuzhnyi2007} Y.V. Kalyuzhnyi, V. Vlachy, P.T. Cummings, Phys.Chem.Lett. 
438 (2007) 238.
\bibitem{Bernard} O. Bernard, L. Blum, J.Chem.Phys. 104 (1996) 4746.
\bibitem{Boublik1970} T. Boublik, J. Chem. Phys. 53 (1970), 471.
\bibitem{Mansoori1971} G. A.  Mansoori, N.F. Carnahan, K.E. Starling, T.W. Leland,
J. Chem. Phys. 54 (1971) 1523.
\bibitem{Kalyuzhnyi1998} Y.V. Kalyuzhnyi, Mol.Phys. 94 (1998) 735
\bibitem{Kalyuzhnyi2000} Y.V. Kalyuzhnyi, V. Vlachy, M.F.Holovko, J. Stat. Phys. 100 (2000) 243.
\bibitem{Shelley1995} J.C. Shelley, G.N. Patey, J. Chem. Phys. 103 (1995) 8299.
\bibitem{Camp1999} P.J. Camp, G.N. Patey, J. Chem. Phys. 111 (1999) 9000.
\bibitem{Daub2003} C.D. Daub, G.N. Patey, P.J. Camp, J. Chem. Phys. 119 (2003) 7952. 
\bibitem{Romero2000} J.M. Romero-Enrique, G. Orkoulas, A.Z. Panagiotopoulos, M.E. Fisher, Phys. Rev. Lett. 85 (2000) 4558.
\bibitem{Stell1989} G. Stell, Y.Q. Zhou, J.Chem.Phys. 91 (1989) 3618.
\bibitem{molsim} J.~Re\v{s}\v{c}i\v{c}, P.~Linse, J.~Comput.~Chem. 36 (2015) 1259.
\bibitem{27} E.W. Lemmon, M.O. McLinden, D.G. Friend, Thermophysical
Properties of Fluid Systems. NIST Chemistry WebBook, NIST
Standard Reference Database Number 69; Linstrom, P. J., Mallard,
W. G., Eds.; National Institute of Standards and Technology, Gaithersburg,
MD, 2011 (http://webbook.nist.gov).
\bibitem{Rebelo2005} L.P.N. Rebelo, J.N.C. Lopes, E. Filipe, J.Phys.Chem.B 109 (2005) 6040.
\bibitem{Shereshevsky1935} I.L. Shereshevsky, J. Phys. Chem. 35 (1935) 1712.
\bibitem{Guggenhein1945} E.A. Guggenhein, J. Chem. Phys. 13 (1945) 253.




\end{thebibliography}
\end{document}